\documentclass[english]{article}
\usepackage[T1]{fontenc}
\usepackage[latin9]{inputenc}
\usepackage{geometry}
\usepackage{float}
\usepackage{amssymb}
\usepackage{esint}

\makeatletter

\providecommand{\tabularnewline}{\\}

\@ifundefined{date}{}{\date{}}
\makeatother

\usepackage{babel}
\begin{document}

\title{Matrix model and Holographic Baryons in the D0-D4 background }

\maketitle
\begin{center}
Si-wen Li\footnote{Email: cloudk@mail.ustc.edu.cn} and Tuo Jia\footnote{Email: jt2011@mail.ustc.edu.cn}
\par\end{center}

\begin{center}
\emph{Department of Modern Physics, }\\
\emph{ University of Science and Technology of China, }\\
\emph{ Hefei 230026, Anhui, China}
\par\end{center}

\vspace{10mm}

\begin{abstract}
We study on the spectrum and short-distance two-body force of holographic
baryons by the matrix model, which is derived from Sakai-Sugimoto
model in D0-D4 background (D0-D4/D8 system). The matrix model is derived
by using the standard technique in string theory and it can describe
multi-baryon system. We re-derive the action of the matrix model from
open string theory on the wrapped baryon vertex, which is embedded
in the D0- D4/D8 system. The matrix model offers a more systematic
approach to the dynamics of the baryons at short distances. In our
system, we find that the matrix model describe stable baryonic states
only if $\zeta=U_{Q_{0}}^{3}/U_{KK}^{3}<2$, where $U_{Q_{0}}^{3}$
is related to the number density of smeared D0-branes. This result
in our paper is exactly the same as some previous presented results
studied on this system as \cite{key-24 Baryons in D0-D4}. We also
compute the baryon spectrum ($k=1$ case) and short-distance two-body
force of baryons ($k=2$ case). The baryon spectrum is modified and
could be able to fit the experimental data if we choose suitable value
for $\zeta$. And the short-distance two-body force of baryons is
also modified by the appearance of smeared D0-branes from the original
Sakai-Sugimoto model. If $\zeta>2$, we find that the baryon spectrum
would be totally complex and an attractive force will appear in the
short-distance interaction of baryons, which may consistently correspond
to the existence of unstable baryonic states.
\end{abstract}
\newpage{}

\section{Introduction}

QCD as the underlying fundamental theory to physicists, has achieved
great successes in particle physics and nuclear physics, however we
still can not describe or predict the behavior of baryons or nucleons
exactly. In high energy physics, it is well-known that nuclear physics
remains one of the most difficult and intriguing branches. The key
problem is that we have not yet understood much about the strong-coupling
QCD since to study on the strong-coupling QCD is hopeless by using
the techniques from perturbative quantum field theory directly. However
about two decades ago, the discovery of AdS/CFT and gauge/string duality
\cite{key-1 Maldacena Large N field,key-2 Maldacena Large N limit,key-3 Witten Ads/Holography,key-4 Ads/CFT User Guide}
became one of the turning points. The AdS/CFT correspondence has been
recognized as a promising framework to understand non-perturbative
aspects of gauge field theory, thus it may be able to provide a new
way to study on nuclear physics. 

There have been many applications or models of AdS/CFT for studying
on strong-coupling QCD such as \cite{key-5 Adding flavor to ads/cft,key-6 DT.Son Ads/CFT hydrodynamics,key-7 DT Son Minkowski correlators,key-8 DT.son QCD and a holographic model of hadrons,key-9 CME from flavored Ads/CFT}.
Several top-down models for QCD inspired by AdS/CFT have also been
proposed in recent years. The most well-known D4-$\mathrm{D}8/\overline{\mathrm{D}8}$
brane system based on Witten's work \cite{key-10 Witten Ads-2} was
built up by Sakai and Sugimoto \cite{key-11 SS model 1,key-12 SS model 2}
(Sakai-Sugimoto model). The Sakai-Sugimoto model consists of $N_{c}$
D4-branes and $N_{f}$ D8-branes as probes. The D4-branes are compactified
on a cycle, offering color numbers, which represent pure Yang-Mills
in the low energy effective theory. Supersymmetry is broken down by
introducing the anti-periodic boundary conditions for fermions on
the compactified cycle. Since only the low energy theory is concerned
in this model, so the dynamics of background geometry produced by
$N_{c}$ D4-branes, can be described by Type II A supergravity on
the large-$N_{c}$ limit. The $N_{f}$ species of massless flavored
quarks are introduced by the embedding $N_{f}$ pairs of probe $\mathrm{D}8/\overline{\mathrm{D}8}$-branes.
The flavor $\mathrm{D}8/\overline{\mathrm{D}8}$-branes are connected
at the IR region of the D4 solitonic solution, which corresponds to
the geometrically broken chiral symmetry in the confined phase for
the dual field theory. In this geometry, the low energy effective
theory of light meson sector comes from the worldvolume theory on
the connected $\mathrm{D}8/\overline{\mathrm{D}8}$-branes. There
is an other solution for the background geometry for this model, i.e.
the D4 black brane solution, which corresponds to the deconfined phase
for the dual field theory \cite{key-13 deconfinement and chiral symmetry}
(However, in fact it is less clear that what phase the deconfined
geometry corresponds to in the dual field theory \cite{key-28 Mandal}).
As it is known that baryons are D-branes wrapped on non-trivial cycles
\cite{key-14 Witten Ads-3,key-15 Holographic nuclear physics,key-23 Gross David},
so in this model, baryons are identified as D4-branes wrapped on a
four-cycle which is named as the ``baryon vertex''. And it has turned
out baryons can be treated as a small instanton configuration in the
worldvolume gauge theory on the probe D8-branes \cite{key-16 Hata Baryons from instantons}.
According to these viewpoints, there are many researches on baryons
or nuclear matters by holography, such as the phase structure \cite{key-15 Holographic nuclear physics,key-29 Siwen,key-30 Cold instanton gas,key-31 Cold nuclear matters}
and the interaction \cite{key-17 SS model nuclear force,key-18 Search for attractive force,key-19 Matrix model for baryons}.
However some results from the Sakai-Sugimoto model is still not realistic
to QCD. One of the most likely reasons may be that the Sakai-Sugimoto
model is a theory with large $N_{c}$ limit, but the real QCD is not.
Therefore some generalizations or modifications of this model have
been proposed such as \cite{key-21 SS model in D0-D4}, and the backreaction
of the flavor branes has also been considered recently as \cite{key-32 Dynamic flavor}.
In our paper, we follow \cite{key-21 SS model in D0-D4} and use the
gauge/gravity duality to study the dynamics of multi-baryons from
D0-D4/D8 system by matrix model proposed in \cite{key-19 Matrix model for baryons}.
By using the matrix model, we calculate some basic properties of holographic
baryons: the spectrum of baryons and the effective two-body potential
both from Sakai-Sugimoto in D0-D4 background as an extension of \cite{key-21 SS model in D0-D4,key-24 Baryons in D0-D4}.

The original matrix model is the effective theory for D0-branes \cite{key-26 BFSS Matrix model},
which can also be understood as a dual description of D=11 supergravity.
As a generalization, the matrix model in \cite{key-19 Matrix model for baryons},
is proposed as the effective theory for baryons or nucleons by holography.
In this matrix model, the rank of the matrix represents the number
of baryons and $k$ baryon branes produce $U(k)$ symmetry for $k$-body
baryons. The diagonal elements of matrices represent the positions
of $k$ baryons while the off-diagonal elements are integrated out.
Besides, the size of baryons are related to the classical values of
a pair of complex $k\times N_{f}$ rectangular matrices, which describe
the dynamics of the strings connecting the flavor branes and the baryon
vertices in low energy effective theory. With all together, it comes
to the well-known Atiyah-Drinfeld-Hitchin-Manin (ADHM) matrix of instantons.
So in our paper, we re-derive the matrix model from Sakai-Sugimoto
in D0-D4 background by using standard technique in string theory.
Such background geometry is produced by $N_{c}$ D4-branes with $N_{0}$
smeared D0-branes. In this background, we follow the original idea
in \cite{key-14 Witten Ads-3,key-23 Gross David}, and recall that
baryons can be identified as D4-branes wrapped on the 4-cycle. Here
we start to use D4'-branes to distinguish such a baryon vertex from
those D4-branes who are responsible for the background geometry. In
the large $N_{c}$ limit, the dynamics of the open strings with both
ends on the D4'-branes and which connects the D4'-branes and the D8-branes
are also relevant, thus the dynamics of $k$ baryons are described
by $U(k)$ gauge theory. And the theory on the D4'-branes is reduced
to a 0+1 dimensional matrix model by considering only the zero modes
along the $S^{4}$ on which the D4'- and the D8-branes are wrapped.
Then it is clear that the matrix model is just the low energy effective
theory for the baryon vertices.

Our motivation for this paper is to study the holographic baryons
by the matrix model derived from Sakai-Sugimoto model in D0-D4 background
(i.e. D0-D4/D8 system). In this paper, since the appearance of smeared
D0-branes could be able to modify the results about baryons from the
original Sakai-Sugimoto model, thus our results may be more close
to the realistic physics. On the other hand, accommodating many bodies
of baryons by such a matrix model would be very easy, so we can even
use this model to describe the interaction among multi-baryons. The
outline of this paper is as follows. In Section II, we give a brief
review of Sakai-Sugimoto model in D0-D4 background (D0-D4/D8 system).
In Section III, we start to derive the matrix model in the D0-D4 background
by considering D4'-D8 gauge theory compactified on $S^{4}$. Since
we keep $k$ baryons close to each other, so only the short-distance
effects are relevant. It turns out that our matrix model can describe
stable baryonic states only for $\zeta=U_{Q_{0}}^{3}/U_{KK}^{3}<2$,
where $U_{Q_{0}}^{3}$ is related to the number density of smeared
D0-branes. And the constraint for stable baryonic states ($\zeta<2$)
is exactly the same as \cite{key-24 Baryons in D0-D4}. And as two
simple examples, we calculate the energy functions of static configurations
with $k=1$ and $k=2$, for one and two flavor(s). In Section IV,
we use our matrix model derived from D0-D4/D8 system to study baryon
spectrum. We determine the size of the holographic baryons with the
case of $k=1$, also for one and two flavor(s). Again it turns out
that the baryon spectrum does make sense only for $\zeta<2$, otherwise
baryons may not exist or be unstable. In Section V, we study the case
of $k=2$ and use the instantons of ADHM matrix as data to calculate
a baryon-baryon potential at short distance for two-flavor case. By
integrating out the auxiliary gauge potential in 0+1 dimension, it
also turns out that there is a universal repulsive core of the two-body
force, but modified by the appearance of smeared D0-branes. And a
short-distance attractive force would appear if $\zeta>2$, which
consistently corresponds to the existence of unstable baryonic states
in two-body system. The summary and conclusion are given in the final
section.

\section{A brief review of Sakai-Sugimoto model in the D0-D4 background}

Here we begin to use D4'-brane to distinguish baryon vertex from those
$N_{c}$ D4-branes which are producing the background geometry. As
we are going to derive the low energy effective theory for D4'-branes
from Sakai-Sugimoto model in D0-D4 background, thus in this section,
we first review the Sakai-Sugimoto model in D0-D4 background geometry
briefly. Some of the results in this section are already presented
in \cite{key-21 SS model in D0-D4,key-24 Baryons in D0-D4}.

\subsection{D0-D4 background geometry}

By taking the near horizon limit, the solution of D4-branes with smeared
D0 charges in Type IIA supergravity reads \cite{key-21 SS model in D0-D4}

\begin{eqnarray}
ds^{2} & = & \left(\frac{U}{R}\right)^{3/2}\left(H_{0}^{1/2}\eta_{\mu\nu}dx^{\mu}dx^{\nu}+H_{0}^{-1/2}f\left(U\right)d\tau^{2}\right)\nonumber \\
 &  & +H_{0}^{1/2}\left(\frac{R}{U}\right)^{3/2}\left(\frac{1}{f\left(U\right)}dU^{2}+U^{2}d\Omega_{4}^{2}\right).\label{eq:D0-D4 metric}
\end{eqnarray}

\noindent We have written this metric (\ref{eq:D0-D4 metric}) in
string frame and $\tau$ is a periodic variable. With the near horizon
limit we also have the formulas for dilaton, Ramond-Ramond (R-R) 4-form
and the 2-form which is

\begin{equation}
e^{\phi}=g_{s}\left(\frac{U}{R}\right)^{3/4}H_{0}^{3/4};\ f_{2}=\frac{A}{U^{4}}\frac{1}{H_{0}^{2}}dU\wedge d\tau;\ f_{4}=dC_{3}=B\epsilon_{4};\label{eq:l-form and dilation}
\end{equation}

\noindent where we have

\begin{equation}
A=\frac{\left(2\pi l_{s}\right)^{7}g_{s}N_{0}}{\omega_{4}V_{4}};\ B=\frac{\left(2\pi l_{s}\right)^{3}N_{c}g_{s}}{\omega_{4}};\ H_{0}=1+\frac{U_{Q_{0}}^{3}}{U^{3}};\ f\left(U\right)=1-\frac{U_{KK}^{3}}{U^{3}},
\end{equation}

\noindent $d\Omega_{4}$, $\epsilon_{4}$ and $\omega_{4}=8\pi^{2}/3$
are the line element, the volume form and the volume of a unit $S^{4}$.
$U_{KK}$ is the coordinate radius of the bottom of the bubble, and
$V_{4}$ is the volume of D4-brane. $N_{0}$ and $N_{c}$ are the
numbers of D0- and D4-branes respectively. D0-branes are smeared in
the $x^{0},...,x^{3}$ directions. The relations to the QCD variables
are deformed as

\begin{equation}
R^{3}=\frac{\lambda l_{s}^{2}}{2M_{KK}};\ g_{s}=\frac{\lambda}{2\pi M_{KK}N_{c}l_{s}};\ U_{KK}=\frac{2}{9}M_{KK}\lambda l_{s}^{2}H_{0}\left(U_{KK}\right).\label{eq:Relations to QCD}
\end{equation}

\noindent Here in order to keep the back reaction of D0-brane, we
also required $N_{0}$ to be of order $N_{c}$ as in \cite{key-25 Liu Hong D instanton}.

\subsection{Embedded D8-branes and baryon vertex in D0-D4 background}

In the Sakai-Sugimoto model, flavors of the dual gauge field theory
are introduced by $N_{f}$ $\mathrm{D}8/\overline{\mathrm{D}8}$-branes
as probes embedded in the background geometry. Here we embed these
$N_{f}$ $\mathrm{D}8/\overline{\mathrm{D}8}$-branes into the background
described by (\ref{eq:D0-D4 metric}) as \cite{key-11 SS model 1,key-21 SS model in D0-D4}.
By taking the probe limit i.e. $N_{c},N_{0}\gg N_{f}$ which makes
the back reaction of the $\mathrm{D}8/\overline{\mathrm{D}8}$-brane
to the background negligible. The fermions created by open strings
can be added to the fundamental representation of $U(N_{c})\times U_{R/L}(N_{f})$
which is treated as groups of chiral symmetry, while the gauge fields
are in adjoint representation.

We employ the viewpoint in Sakai-Sugimoto model, baryons have been
provided as a D4'-brane wrapped on $S^{4}$ which is called the baryon
vertex \cite{key-14 Witten Ads-3,key-15 Holographic nuclear physics,key-23 Gross David}.
Such D4'-branes have to attach the ends of $N_{c}$ fundamental strings
since the $S^{4}$ is supported by $N_{c}$ units of a R-R flux in
the supergravity solution. In this way, the baryon charge equals to
$N_{c}$ quark charge in the corresponding field theory. According
to these arguments, baryons are also the D4'-branes wrapped on $S^{4}$
in our D0-D4 background geometry (\ref{eq:D0-D4 metric}). As we are
going to study on the baryons, thus we will focus on the baryon vertex
in the Sakai-Sugimoto model in D0-D4 background in next sections.

\section{The matrix model from D0-D4/D8 system}

In this section, we derive the matrix quantum mechanics and we are
going to use the action to study the baryon spectrum and the two-body
interaction by an effective potential. As mentioned, the matrix model
is just the low energy effective action on the $k$ D4'-branes embedded
in flavor $N_{f}$ D8-branes in the geometry described by (\ref{eq:D0-D4 metric}).
Here we first give our result i.e. the action of our matrix model
in D0-D4/D8 system, and we leave the details for derivation in next
two subsections.

\subsection{The action}

The action of our matrix model for baryons from D0-D4/D8 system is 

\begin{eqnarray}
S & = & \frac{\lambda N_{c}M_{KK}}{54\pi}\left(1+\zeta\right)^{3/2}\mathrm{Tr}\int dt\bigg[\left(D_{0}X^{M}\right)^{2}-\frac{2}{3}\left(1-\frac{1}{2}\zeta\right)M_{KK}^{2}\left(X^{4}\right)^{2}\nonumber \\
 &  & +D_{0}\bar{\omega}_{i}^{\dot{\alpha}}D_{0}\omega_{i\dot{\alpha}}-\frac{1}{6}\left(1-\frac{1}{2}\zeta\right)M_{KK}^{2}\bar{\omega}_{i}^{\dot{\alpha}}\omega_{i\dot{\alpha}}\nonumber \\
 &  & +\frac{3^{6}\pi^{2}}{4\lambda^{2}M_{KK}^{4}}\frac{1}{\left(1+\zeta\right)^{4}}\left(\vec{D}\right)^{2}+\vec{D}\cdot\vec{\tau}_{\ \dot{\beta}}^{\dot{\alpha}}\bar{X}^{\dot{\beta}\alpha}X_{\dot{\alpha}\alpha}+\vec{D}\cdot\vec{\tau}_{\ \dot{\beta}}^{\dot{\alpha}}\bar{\omega}^{\dot{\beta}\alpha}\omega_{\dot{\alpha}\alpha}\bigg]\nonumber \\
 &  & +N_{c}\mathrm{Tr}\int dtA_{0}\ .\label{eq:Matrix action}
\end{eqnarray}

\noindent As we can see, (\ref{eq:Matrix action}) describes a quantum
mechanical system with $U\left(k\right)$ symmetry where $k$ is the
baryon number and the trace is taken over $U\left(k\right)$ adjoint
representation. Here $M=1,2,3,4$ and $\lambda=g_{YM}^{2}N_{c}$ is
the 't Hooft coupling. The unique dimension-ful constant $M_{KK}$
is defined as (\ref{eq:Relations to QCD}). $\vec{D}$ and $A_{0}$
are auxiliary fields while the fields $X^{M}$ and $\omega$ are the
dynamical fields. $\zeta$ is defined as $\zeta=U_{Q_{0}}^{3}/U_{KK}^{3}$.
Note that all the fields are bosonic and this matrix model describes
the $k$-baryon system in D0-D4/D8 system according to the holographic
principle.

The matrix model (\ref{eq:Matrix action}) is a deformed matrix model
of \cite{key-19 Matrix model for baryons}. Without smeared D0-branes,
i.e. setting $\zeta=0$, (\ref{eq:Matrix action}) returns back to
the matrix model in \cite{key-19 Matrix model for baryons}. Therefore
the matrix model (\ref{eq:Matrix action}) is also a deformed ADHM
matrix model as claimed in \cite{key-19 Matrix model for baryons}.
By integrating out of the auxiliary field $\vec{D}$, it yields a
potential of a commutator term as $\left(\mathrm{Tr}\left[X,X\right]\right)^{2}$.
So our matrix model looks also close to the BFSS Matrix theory \cite{key-26 BFSS Matrix model}
(which is understood as an effective description of M-theory) or the
IKKT matrix model \cite{key-27 Ishibashi N} if the last term and
the mass term are absent. Note that the quadratic term of $X^{4}$
and $\omega$ would be negative if $\zeta>2$ which corresponds to
a system with imaginary mass. It may be understood as the constraint
for the stable states of baryons in D0-D4 system i.e. if $\zeta>2$,
baryons may not be stable in this system. This viewpoint is exactly
the same as \cite{key-24 Baryons in D0-D4} by using the instanton
views for baryons. Since we will use our matrix model to describe
baryons in D0-D4/D8 system, thus only $\zeta<2$ is considered here
for stable baryons. 

As low energy effective theory, all the fields in our matrix model
(\ref{eq:Matrix action}) is similar to the matrix model of \cite{key-19 Matrix model for baryons}.
Therefore we employ the symmetry for our matrix model from \cite{key-19 Matrix model for baryons}.
And the representation of the fields is summarized in Table 1. The
covariant derivative is defined as

\begin{eqnarray}
D_{0}X^{M} & = & \partial_{0}X^{M}-i\left[A_{0},X^{M}\right],\nonumber \\
D_{0}\omega=\partial_{0}\omega-iA_{0}\omega & ; & D_{0}\bar{\omega}=\partial_{0}\bar{\omega}+iA_{0}\bar{\omega}.
\end{eqnarray}

\noindent So the matrix model (\ref{eq:Matrix action}) is a quantum
mechanical system with the following symmetry

\[
U\left(k\right)\times SU\left(N_{f}\right)\times SO\left(3\right),
\]

\noindent $N_{f}$ is the number of the flavors in QCD. The first
symmetry $U\left(k\right)$ is a local symmetry while the symmetry
$SU\left(N_{f}\right)\times SO\left(3\right)$ is a global symmetry.
$SO\left(3\right)$ could be interpreted as the rotation symmetry
in our space where the baryons live. If we embed the rotation symmetry
as

\[
SO\left(3\right)\subset SO\left(4\right)\simeq SU\left(2\right)\times SU\left(2\right),
\]

\noindent then the symmetry of the action would be easier to understand.
And the additional dimension corresponds to the holographic dimension.
By the mass deformation, the $SO\left(4\right)$ symmetry breaks down
to the $SO\left(3\right)$ symmetry.

\noindent 
\begin{table}
\noindent \begin{centering}
\begin{tabular}{|c|c|c|c|c|}
\hline 
Fields & index & $U\left(k\right)$ & $SU\left(N_{f}\right)$ & $SU\left(2\right)\times SU\left(2\right)$\tabularnewline
\hline 
\hline 
$X^{M}$ & $M=1,2,3,4$ & adj & 1 & (2,2)\tabularnewline
\hline 
$\omega_{i\dot{\alpha}}$ & $\dot{\alpha}=1,2$;$i=1,2...N_{f}$ & adj & fund & (1,2)\tabularnewline
\hline 
$A_{0}$ &  & adj & 1 & (1,1)\tabularnewline
\hline 
$D_{s}$ & $s=1,2,3$ & 1 & 1 & (1,3)\tabularnewline
\hline 
\end{tabular}
\par\end{centering}

\protect\caption{Fields in the matrix model}
\end{table}

\subsection{Derivation from Holography}

In D0-D4/D8 system, our concern is the D4'-branes wrapped on $S^{4}$
in the background (\ref{eq:D0-D4 metric}), whose low energy effective
theory is described by the matrix model (\ref{eq:Matrix action}).
In AdS/CFT duality, such a D4'-brane is named as the ``baryon vertex''
\cite{key-14 Witten Ads-3}, which is responsible for creating or
annihilating baryonic states in the dual field theory. 

Since the D4'-branes live inside the flavor branes, the action of
the D4'-brane is related to the background geometry, the R-R flux
and also affected by the presence of the flavor branes. So in the
low energy effective theory of baryonic D4'-branes, the strings connecting
the baryon vertices and the flavor branes provide the field $\omega$
in the bi-fundamental representation. The location of the D4'-branes
in the transverse directions is specified by the diagonal eigenvalues
of the field $X^{M}$ in the adjoint representation. Thus these diagonal
eigenvalues represent the locations of the baryons in our real 3-dimensional
space. Due to the curved geometry and the flux, only the bosonic fields
are kept here since the deformation between our matrix model and ADHM
matrix model breaks the supersymmetry explicitly. So we will not care
about the fermion part in our theory.

As in \cite{key-19 Matrix model for baryons}, we assume that the
$S^{4}$ dependence can be trivially reduced although the D4'-branes
are inside the flavor branes and wrapped on $S^{4}$, that means a
dimensional reduction with no dependence along $S^{4}$. Therefore
the derived action is in the time dimension only i.e. depended on
time only. So next, we will derive the low energy effective action
of $k$ brane vertices system by using standard technique in string
theory.

\subsubsection{Derivation from DBI part}

Let us start from the action for a single D4'-brane which is

\begin{eqnarray}
S_{D4'} & = & S_{DBI}+S_{CS}\ ,\nonumber \\
S_{DBI} & = & -T_{D4}\int d^{5}\xi e^{-\phi}\sqrt{-\det\left(G_{MN}+2\pi\alpha^{\prime}F_{MN}\right)},\nonumber \\
S_{CS} & = & \frac{1}{2\pi}\int C_{3}\wedge F_{2}.\label{eq:D-brane action}
\end{eqnarray}

\noindent Here we have used $\xi$ to represent the coordinates on
baryon vertex. The convenient coordinates are used here as in \cite{key-11 SS model 1,key-21 SS model in D0-D4}

\begin{eqnarray}
y=r\cos\theta & ; & Z=r\sin\theta,\nonumber \\
U^{3}=U_{KK}^{3}+U_{KK}r^{2} & ; & \theta=\frac{3}{2}\frac{U_{KK}^{1/2}}{R^{3/2}H_{0}^{1/2}\left(U_{KK}\right)}\tau,
\end{eqnarray}

\noindent where the flavor D8-brane is located at $y=0$. We consider
a stable D4'-brane situated at $r=0$ wrapped on $S^{4}$, in this
case the DBI action is 

\begin{equation}
S_{DBI}=-\frac{T_{D4'}\omega_{4}}{g_{s}}\int dtH_{0}^{1/4}UR^{3}\left(R/U\right)^{3/4}\sqrt{-G_{00}},
\end{equation}

\noindent where the induced metric is

\begin{equation}
G_{00}=\left(\frac{U}{R}\right)^{3/2}H_{0}^{1/2}\left[-1+\left(\partial_{0}X^{i}\right)^{2}\right]+\frac{4}{9}\frac{U_{KK}}{U}\left(\frac{R}{U}\right)^{3/2}H_{0}^{1/2}\left(\partial_{0}Z\right)^{2}.
\end{equation}

\noindent We have induced the metric on the worldvolume of D4'-branes
to the low energy effective theory and the index of $X^{i}$ is $i=1,2,3$,
i.e. runs for 3-dimensional space, so we obtain

\begin{equation}
S_{DBI}=-\frac{T_{D4'}\omega_{4}}{g_{s}}\int dtH_{0}^{1/2}UR^{3}\sqrt{1-\left(\partial_{0}X^{i}\right)^{2}-\frac{4}{9}\frac{U_{KK}}{U}\left(\frac{R}{U}\right)^{3}\left(\partial_{0}Z\right)^{2}}.\label{eq:DBI in Z coordinate}
\end{equation}

\noindent Since we keep $k$ baryons at short distance and only the
low energy effective theory is considered here, so we just need to
expand (\ref{eq:DBI in Z coordinate}) for small $Z$ and small $X$
and define

\begin{eqnarray}
X^{4} & = & \frac{2}{3}\left(\frac{R}{U_{KK}}\right)^{3/2}Z\ ,\nonumber \\
\zeta & = & \frac{U_{Q_{0}}^{3}}{U_{KK}^{3}}\ ,
\end{eqnarray}

\noindent yields a quadratic action

\begin{equation}
S_{DBI}=\frac{\lambda N_{c}M_{KK}}{27\pi}\left(1+\zeta\right)^{3/2}\int dt\left[-1+\frac{1}{2}\left(\partial_{0}X^{M}\right)^{2}-\frac{1}{6}\left(\zeta-2\right)M_{KK}^{2}\left(X^{4}\right)^{2}\right].\label{eq:Matrix from DBI}
\end{equation}

\noindent The high orders of $X,\ Z$ and their derivatives have been
dropped off.

The kinetic term and mass term for $X$ in matrix action (\ref{eq:Matrix action})
is given by (\ref{eq:Matrix from DBI}). However this leaves the mass
term for $\omega$. We have assumed that the mass term for $\omega$
is

\begin{equation}
-\frac{1}{12}\left(\zeta-2\right)M_{KK}^{2}\bar{\omega}_{i}^{\dot{\alpha}}\omega_{i\dot{\alpha}}.\label{eq:mass term for w}
\end{equation}

\noindent This (\ref{eq:mass term for w}) is a natural guess from
a comparison with  \cite{key-19 Matrix model for baryons}. Since
our matrix model is the deformation from the model in \cite{key-19 Matrix model for baryons},
we simply set the mass term of $\omega$ as $1/4$ times of the mass
term of $X^{4}$ which is an assumption from \cite{key-19 Matrix model for baryons}.

\subsubsection{Commutator terms }

We will compute the commutator terms in the matrix action (\ref{eq:Matrix action}).
First we expand the generic Dirac-Born-Infield (DBI) action of a Dp-brane
(\ref{eq:D-brane action}) to the quadratic order

\begin{equation}
S_{DBI}\backsimeq\left(2\pi\alpha^{\prime}\right)^{2}\frac{1}{4}T_{Dp}\int d^{5}\xi e^{-\phi}\sqrt{-\det G_{MN}}G_{MN}G_{PQ}F^{MP}F^{NQ}\ .
\end{equation}

\noindent The relevant relations from T-duality here is $\left(2\pi\alpha^{\prime}\right)A_{M}=X^{N}G_{NM}$
. We therefore have

\begin{eqnarray}
\left(2\pi\alpha^{\prime}\right)^{2}G_{MN}G_{PQ}F^{MP}F^{NQ} & = & 2G^{00}G_{ij}D_{0}X^{i}D_{0}X^{j}+2G^{00}G_{zz}D_{0}ZD_{0}Z\nonumber \\
 &  & -\frac{1}{\left(2\pi\alpha^{\prime}\right)^{2}}\left[X^{i},X^{j}\right]\left[X^{k},X^{l}\right]G_{ik}G_{jl}\nonumber \\
 &  & -\frac{2}{\left(2\pi\alpha^{\prime}\right)^{2}}\left[X^{i},Z\right]\left[X^{j},Z\right]G_{ij}G_{zz}\nonumber \\
 & = & -2\left(D_{0}X^{M}\right)^{2}-\frac{4}{3^{6}\pi^{2}}\left(1+\zeta\right)^{4}\lambda^{2}M_{KK}^{4}\left[X^{M},X^{N}\right]^{2}\ .\label{eq:commutator terms}
\end{eqnarray}

\noindent We have used the value of the metric at $U=U_{KK}$. The
commutator term can be rewritten by an auxiliary field $\overrightarrow{D}$
if we consider the following action

\begin{equation}
S=c\int dt\mathrm{Tr}\left[2\left(2\pi\alpha^{\prime}\right)^{2}\vec{D}^{2}+\vec{D}\cdot\vec{\tau}_{\ \dot{\beta}}^{\dot{\alpha}}\bar{a}^{\prime\dot{\beta}\alpha}a_{\alpha\dot{\alpha}}^{\prime}\right]\ .
\end{equation}

\noindent Integrating out the field $\vec{D}$, yields 

\begin{equation}
S=c\int dt\mathrm{Tr}\left[\frac{1}{16\pi^{2}\alpha^{\prime2}}\left[a_{m}^{\prime},a_{n}^{\prime}\right]^{2}\right]\ .\label{eq:action with auxiliary field}
\end{equation}

\noindent Comparing (\ref{eq:action with auxiliary field}) with (\ref{eq:Matrix from DBI}),
we have

\begin{equation}
S=\frac{\lambda N_{c}M_{KK}}{54\pi}\left(1+\zeta\right)^{3/2}\int dt\left[\frac{3^{6}\pi^{2}}{4\lambda^{2}M_{KK}^{4}}\frac{1}{\left(1+\zeta\right)^{4}}\left(\vec{D}\right)^{2}+\vec{D}\cdot\vec{\tau}_{\ \dot{\beta}}^{\dot{\alpha}}\bar{X}^{\beta\alpha}X_{\alpha\dot{\alpha}}\right]\ ,\label{eq:action with auxiliary field -2}
\end{equation}

\subsubsection{Chern-Simons term}

Finally, let us consider the Chern-Simons part of the Dp-brane action.
Since $f_{4}=dC_{3}$ is given in (\ref{eq:l-form and dilation}),
thus the Chern-Simons term in (\ref{eq:D-brane action}) can be rewritten
as

\begin{eqnarray}
S_{CS} & = & \frac{1}{2\pi}\int C_{3}\wedge F_{2}\nonumber \\
 & = & \frac{1}{2\pi}\frac{1}{2\cdot3!}\mathrm{Tr}\int d^{5}\xi\epsilon^{\mu_{1}\mu_{2}\mu_{3}\alpha\beta}C_{\mu_{1}\mu_{2}\mu_{3}}F_{\alpha\beta}\nonumber \\
 & = & N_{c}\mathrm{Tr}\int dtA_{0},\label{eq:WZ term}
\end{eqnarray}

\noindent (\ref{eq:WZ term}) is nothing but the last term in matrix
action (\ref{eq:Matrix from DBI}).

\section{\noindent Baryon spectrum}

In this section, we will use our matrix model (\ref{eq:Matrix action})
to study baryon spectrum of D0-D4/D8 system for $k=1$ case, i.e the
single baryon case. So first let us compute the Hamiltonian for a
single baryon $k=1$ with general $N_{f}$ first. Since for $k=1$
the field $X$ is a number, thus all commutators with $X$ could be
dropped. This leaves the terms of $\omega$.

\begin{eqnarray}
S_{\vec{D}} & = & \frac{\lambda N_{c}M_{KK}}{54\pi}\left(1+\zeta\right)^{3/2}\int dt\left[\frac{3^{6}\pi^{2}}{4\lambda^{2}M_{KK}^{4}}\frac{1}{\left(1+\zeta\right)^{4}}\left(\vec{D}\right)^{2}+\vec{D}\cdot\vec{\tau}_{\ \dot{\beta}}^{\dot{\alpha}}\bar{\omega}^{\dot{\beta}\alpha}\omega_{\dot{\alpha}\alpha}\right]\nonumber \\
 & = & -\frac{\lambda N_{c}M_{KK}}{54\pi}\left(1+\zeta\right)^{3/2}\int dt\frac{\lambda^{2}M_{KK}^{4}}{3^{6}\pi^{2}}\left(1+\zeta\right)^{4}\bigg[4\omega_{1}^{i}\omega_{2}^{i*}\omega_{2}^{j}\omega_{1}^{j*}+\left(\omega_{1}^{i}\omega_{1}^{i*}\right)^{2}\nonumber \\
 &  & +\left(\omega_{2}^{i}\omega_{2}^{i*}\right)^{2}-2\omega_{1}^{i}\omega_{1}^{i*}\omega_{2}^{i}\omega_{2}^{i*}\bigg]\ .\label{eq:ADHM potential}
\end{eqnarray}

\noindent The so-called ADHM potential is given by (\ref{eq:ADHM potential}).
Basically we should solve it for the construction of instantons in
the ADHM formalism, however it is equivalent to minimize the ADHM
potential here. And since there is no dynamics for $A_{0}$ in our
matrix action (\ref{eq:Matrix action}), thus we can integrate it
out and the terms including $A_{0}$ become

\begin{eqnarray}
S_{A_{0}} & = & \frac{\lambda N_{c}M_{KK}}{54\pi}\left(1+\zeta\right)^{3/2}\nonumber \\
 &  & \times\int dt\left[iA_{0}\bar{\omega}_{i}^{\dot{\alpha}}\partial_{0}\omega_{\dot{\alpha}i}-i\partial_{0}\bar{\omega}_{i}^{\dot{\alpha}}A_{0}\omega_{\dot{\alpha}i}+\left(A_{0}\right)^{2}\bar{\omega}_{i}^{\dot{\alpha}}\omega_{\dot{\alpha}i}+\frac{54\pi}{\lambda M_{KK}}\frac{1}{\left(1+\zeta\right)^{3/2}}A_{0}\right]\ .\label{eq:action part of A0}
\end{eqnarray}

\noindent So the equation of motion for $A_{0}$ is

\begin{equation}
i\bar{\omega}_{i}^{\dot{\alpha}}\partial_{0}\omega_{\dot{\alpha}i}-i\partial_{0}\bar{\omega}_{i}^{\dot{\alpha}}\omega_{\dot{\alpha}i}+2A_{0}\bar{\omega}_{i}^{\dot{\alpha}}\omega_{\dot{\alpha}i}+\frac{54\pi}{\lambda M_{KK}}\frac{1}{\left(1+\zeta\right)^{3/2}}=0.\label{eq:EOM for A0}
\end{equation}

\noindent By inserting (\ref{eq:EOM for A0}) to (\ref{eq:action part of A0})
we obtain

\begin{equation}
S_{A_{0}}=\frac{\lambda N_{c}M_{KK}}{54\pi}\left(1+\zeta\right)^{3/2}\int dt\left[-\frac{1}{4\bar{\omega}_{i}^{\dot{\alpha}}\omega_{\dot{\alpha}i}}\left(i\bar{\omega}_{i}^{\dot{\alpha}}\partial_{0}\omega_{\dot{\alpha}i}-i\partial_{0}\bar{\omega}_{i}^{\dot{\alpha}}\omega_{\dot{\alpha}i}+\frac{54\pi}{\lambda M_{KK}}\frac{1}{\left(1+\zeta\right)^{3/2}}\right)^{2}\right]\ .
\end{equation}

\noindent Then we use the definition of the momentum conjugate to
the field $\omega$

\begin{eqnarray}
P_{i}^{\dot{\alpha}} & = & \frac{\partial S}{\partial\left(\partial_{0}\omega_{\dot{\alpha}}^{i}\right)}\nonumber \\
 & = & \frac{\lambda N_{c}M_{KK}}{54\pi}\left(1+\zeta\right)^{3/2}\left[\partial_{0}\bar{\omega}_{i}^{\dot{\alpha}}-\frac{1}{2\bar{\omega}_{j}^{\dot{\gamma}}\omega_{\dot{\gamma}j}}\left(i\bar{\omega}_{k}^{\dot{\beta}}\partial_{0}\omega_{\dot{\beta}k}-i\partial_{0}\bar{\omega}_{k}^{\dot{\beta}}\omega_{\dot{\beta}k}+\frac{54\pi}{\lambda M_{KK}}\frac{1}{\left(1+\zeta\right)^{3/2}}\right)i\bar{\omega}_{i}^{\dot{\alpha}}\right]\ .\label{eq:conjugate momentum}
\end{eqnarray}

\noindent Therefore we obtain the Hamiltonian

\begin{eqnarray}
H & = & P_{i}^{\dot{\alpha}}\partial_{0}\omega_{i}^{\dot{\alpha}}+\bar{P}_{i}^{\dot{\alpha}}\partial_{0}\bar{\omega}_{i}^{\dot{\alpha}}-L\nonumber \\
 & = & \frac{\lambda N_{c}M_{KK}}{54\pi}\left(1+\zeta\right)^{3/2}\bigg[\partial_{0}\bar{\omega}_{i}^{\dot{\alpha}}\partial_{0}\omega_{i}^{\dot{\alpha}}+\frac{1}{6}\left(1-\frac{1}{2}\zeta\right)M_{KK}^{2}\bar{\omega}_{i}^{\dot{\alpha}}\omega_{i\dot{\alpha}}\nonumber \\
 &  & +\frac{\lambda^{2}M_{KK}^{4}}{3^{6}\pi^{2}}\left(1+\zeta\right)^{4}\left(4\omega_{1}^{i}\omega_{2}^{i*}\omega_{2}^{j}\omega_{1}^{j*}+\left(\omega_{1}^{i}\omega_{1}^{i*}\right)^{2}+\left(\omega_{2}^{i}\omega_{2}^{i*}\right)^{2}-2\omega_{1}^{i}\omega_{1}^{i*}\omega_{2}^{i}\omega_{2}^{i*}\right)\nonumber \\
 &  & +\frac{1}{4\bar{\omega}_{i}^{\dot{\alpha}}\omega_{\dot{\alpha}i}}\left(\left(\frac{54\pi}{\lambda M_{KK}}\right)^{2}\frac{1}{\left(1+\zeta\right)^{3}}+\left(\bar{\omega}_{i}^{\dot{\alpha}}\partial_{0}\omega_{\dot{\alpha}i}-\partial_{0}\bar{\omega}_{i}^{\dot{\alpha}}\omega_{\dot{\alpha}i}\right)^{2}\right)\bigg]\ .\label{eq:Hamiltonian single baryon}
\end{eqnarray}

\subsection{Single flavor}

For single flavor case, i.e. $N_{f}=1$, we use the following ansatz
for $\omega$

\begin{equation}
\omega_{\dot{\alpha}}=\rho_{\dot{\alpha}};\ \bar{\omega}_{\dot{\alpha}}=\rho_{\dot{\alpha}}^{*}.\label{eq:ansatz for single flavor}
\end{equation}

\noindent Then the Hamiltonian (\ref{eq:Hamiltonian single baryon})
is

\begin{eqnarray}
H & = & \frac{\lambda N_{c}M_{KK}}{54\pi}\left(1+\zeta\right)^{3/2}\nonumber \\
 &  & \times\left[\frac{1}{2\rho^{2}}\left(\frac{27\pi}{\lambda M_{KK}}\right)^{2}\frac{1}{\left(1+\zeta\right)^{3}}+\frac{1}{3}\left(1-\frac{1}{2}\zeta\right)M_{KK}^{2}\rho^{2}+\frac{4\lambda^{2}M_{KK}^{4}}{3^{6}\pi^{2}}\left(1+\zeta\right)^{4}\rho^{4}\right]\ ,\label{eq:Hamiltonian of rho single flavor}
\end{eqnarray}

\noindent where we have used the definition $2\rho^{2}=\rho_{1}\rho_{1}^{*}+\rho_{2}\rho_{2}^{*}$.
Let us analyze the terms in (\ref{eq:Hamiltonian of rho single flavor}).
In the soliton picture, the first term can be interpreted as a self-repulsion
of the instanton which is induced by the Chern-Simons term with the
path-integration of $A_{0}$. The second term comes from the mass
term of the matrix model and the curved background geometry while
the third term comes from the path-integration over auxiliary field
$\vec{D}$, which corresponds to the ADHM potential. Thus all the
terms are physical and modified by the appearance of smeared D0-branes.

It is easy to find that the Hamiltonian could be minimized by a nonzero
$\rho$. In the large $\lambda$ limit, we can reduce the Hamiltonian
(\ref{eq:Hamiltonian of rho single flavor}) to a linear formula by
putting $\rho=x\lambda^{\alpha}M_{KK}^{-1}$. Here both $x$ and $\alpha$
are constants. And similar as done in \cite{key-19 Matrix model for baryons},
every term in (\ref{eq:Hamiltonian of rho single flavor}) scales
with large $\lambda$ limit. As a result, the second term in (\ref{eq:Hamiltonian of rho single flavor})
can be negligible, then the value for $\rho$ to minimize this Hamiltonian
is computed as

\begin{equation}
\rho=2^{-2/3}3^{2}\pi^{2/3}\lambda^{-2/3}M_{KK}^{-1}\left(1+\zeta\right)^{-7/6},
\end{equation}

\noindent and 

\begin{equation}
H_{min}=2^{-5/3}\pi^{-1/3}\lambda^{1/3}N_{c}M_{KK}\left(1+\zeta\right)^{5/6}.
\end{equation}

\subsection{Two flavors}

In this subsection, we need to focus on a more realistic case which
has two flavors. In order to eliminate the contribution from the ADHM
potential, we first have to satisfy the ADHM constraints $\vec{\tau}_{\ \dot{\beta}}^{\dot{\alpha}}\bar{\omega}_{i}^{\dot{\beta}}\omega_{\dot{\alpha i}}=0$,
or equivalently, use the following ansatz

\begin{equation}
\sum_{i=1}^{N_{f}}\omega_{\dot{\alpha}=1}^{i}\omega_{\dot{\alpha}=2}^{i*}=\sum_{i=1}^{N_{f}}\omega_{\dot{\alpha}=2}^{i}\omega_{\dot{\alpha}=1}^{i*}=0;\ \sum_{i=1}^{N_{f}}\left|\omega_{\dot{\alpha}=1}^{i}\right|^{2}=\sum_{i=1}^{N_{f}}\left|\omega_{\dot{\alpha}=2}^{i}\right|^{2}\ .
\end{equation}

\noindent The ADHM potential disappears if this condition is satisfied.
And without loss of generality, this can be achieved by using the
following choice 

\begin{equation}
\omega_{\dot{\alpha}}^{i=1}=\left(\begin{array}{c}
\rho\\
0
\end{array}\right)_{\dot{\alpha}};\ \omega_{\dot{\alpha}}^{i=2}=\left(\begin{array}{c}
0\\
\rho
\end{array}\right)_{\dot{\alpha}}.
\end{equation}

\noindent After including $X^{4}$-dependence, the Hamiltonian is

\begin{eqnarray}
H & = & \frac{\lambda N_{c}M_{KK}}{54\pi}\left(1+\zeta\right)^{3/2}\bigg[\frac{1}{2\rho^{2}}\left(\frac{27\pi}{\lambda M_{KK}}\right)^{2}\frac{1}{\left(1+\zeta\right)^{3}}+\frac{1}{3}\left(1-\frac{1}{2}\zeta\right)M_{KK}^{2}\rho^{2}\nonumber \\
 &  & +\frac{2}{3}\left(1-\frac{1}{2}\zeta\right)M_{KK}^{2}\left(X^{4}\right)^{2}\bigg]\ .\label{eq:Hamiltonian finite Nc two flavor}
\end{eqnarray}

\noindent This can be minimized at

\begin{equation}
\rho=2^{-1/4}3^{7/4}\pi^{1/2}\lambda^{-1/2}\left(1-\frac{1}{2}\zeta\right)^{-1/4}\left(1+\zeta\right)^{-3/4}M_{KK}^{-1}\ .
\end{equation}

\noindent Then the minimum value of the Hamiltonian is

\begin{equation}
H_{min}=\frac{N_{c}M_{KK}}{\sqrt{6}}\sqrt{1-\frac{1}{2}\zeta}\ .\label{eq:Hmin 2 flavor}
\end{equation}

\noindent In the soliton approach \cite{key-16 Hata Baryons from instantons,key-24 Baryons in D0-D4},
the value of $\rho$ is the size of instantons. Obviously in our theory,
we find that all the values are modified by $\zeta$ which is related
to the appearance of smeared D0-branes and (\ref{eq:Hmin 2 flavor})
would be totally complex if $\zeta>2$.

\subsection{Quantization}

Since we have obtained the results on the vacuum of the matrix model,
we can then quantize the Hamiltonian for the $k=1,\ N_{f}=2$ circumstance
and the results should correspond to the baryon spectrum. We first
rewrite the Hamiltonian used in \cite{key-16 Hata Baryons from instantons}
since we are going to use the same tricks, so we have

\begin{equation}
H=\frac{\lambda N_{c}M_{KK}}{54\pi}\left[\frac{2}{5\rho^{2}}\left(\frac{27\pi}{\lambda M_{KK}}\right)^{2}+\frac{1}{3}M_{KK}^{2}\rho^{2}+\frac{2}{3}M_{KK}^{2}\left(X^{4}\right)^{2}\right]\ .\label{eq:Hamiltonian in Ref.}
\end{equation}

\noindent By comparing (\ref{eq:Hamiltonian in Ref.}) with (\ref{eq:Hamiltonian finite Nc two flavor}),
the energy spectrum in our system can be obtained easily by replacing
the $Q$ and $\omega_{\rho},\omega_{Z}$ used in \cite{key-16 Hata Baryons from instantons}
as

\noindent 
\begin{equation}
Q\rightarrow\frac{5}{4}Q\left(1+\zeta\right)^{-3/2};\ \left(\omega_{\rho}\ \mathrm{or}\ \omega_{Z}\right)\rightarrow\left(1+\zeta\right)^{3/4}\left(1-\frac{1}{2}\zeta\right)^{1/2}\left(\omega_{\rho}\ \mathrm{or}\ \omega_{Z}\right)\ .
\end{equation}

\noindent Then the mass formula for the baryon excitation is

\begin{equation}
M=M_{0}+\left(1+\zeta\right)^{3/4}\left(1-\frac{1}{2}\zeta\right)^{1/2}\left[\sqrt{\frac{\left(l+1\right)^{2}}{6}+\frac{N_{c}^{2}}{6}\left(1+\zeta\right)^{-3/2}}+\frac{2\left(n_{\rho}+n_{z}\right)+2}{\sqrt{6}}\right]\ ,\label{eq:Baryon mass}
\end{equation}

\noindent where $M_{0}=\frac{\lambda N_{c}M_{KK}}{27\pi}$. In order
to fit the data from experimental values, we set $\zeta=1.9303$ with
$N_{c}=3$ for real QCD and set $M_{KK}=945\mathrm{MeV}$ to fit the
mass of $\rho$ meson. The baryon mass spectrum from our matrix model
(\ref{eq:Baryon mass}) is listed in Table 2 which is more close to
the experimental data . As a comparison, we also list experimental
data in Table 3 and the superscripts $\pm$ represent the parity.
The subscript $*$ here is used to indicate that evidence for the
existence of the baryonic states is poor.

Such a baryon spectrum has already been obtained in \cite{key-16 Hata Baryons from instantons}.
However their original results are much larger than the experimental
data if fitting the experiment data by setting $M_{KK}=945\mathrm{MeV}$
(mass of $\rho$ meson) with $N_{c}=3$. And in \cite{key-19 Matrix model for baryons},
the baryon spectrum would be still larger than the experimental data
if fitting the experiment with the same value for $M_{KK}$ and $N_{c}$.
Most likely, the reason is that Sakai-Sugimoto model describes the
QCD with large $N_{c}$ limit by holography, but the real QCD is a
theory with $N_{c}=3$. Therefore, in our D0-D4/D8 system, we suggest
to give an effective description for $N_{c}=3$ QCD by adjusting the
number density of D0-branes i.e. the parameter $\zeta$ in our system.
Note that our result (\ref{eq:Baryon mass}) does not make sense if
$\zeta>2$ since the mass spectrum would be totally imaginary. As
mentioned, the stable baryonic state may not exist if $\zeta>2$ in
D0-D4/D8 system, which would be quite different from the original
Sakai-Sugimoto model.

\begin{table}[H]
\begin{centering}
\begin{tabular}{|c|c|c|c|c|}
\hline 
$\left(n_{\rho},n_{z}\right)$ & $\left(0,0\right)$ & $\left(1,0\right)\ \left(0,1\right)$ & $\left(1,1\right)$ $\left(2,0\right)/\left(0,2\right)$ & $\left(2,1\right)/\left(0,3\right)$ $\left(1,2\right)/\left(3,0\right)$\tabularnewline
\hline 
\hline 
$\begin{array}{c}
N\left(l=1\right)\\
\Delta\left(l=3\right)
\end{array}$ & $\begin{array}{c}
945^{+}\\
1237^{+}
\end{array}$ & $\begin{array}{c}
1268^{+}\\
1560^{+}
\end{array}$ $\begin{array}{c}
1268^{-}\\
1560^{-}
\end{array}$ & $\begin{array}{c}
1590^{-}\\
1882^{-}
\end{array}$ $\begin{array}{c}
1590^{+},1590^{+}\\
1882^{+},1882^{+}
\end{array}$ & $\begin{array}{c}
1913^{-},1913^{-}\\
2205^{-},2205^{-}
\end{array}$$\begin{array}{c}
1913^{+},1913^{+}\\
2205^{+},2205^{+}
\end{array}$\tabularnewline
\hline 
\end{tabular}
\par\end{centering}

\protect\caption{Baryon spectrum of mass from (\ref{eq:Baryon mass})}
\end{table}

\begin{table}[H]
\begin{centering}
\begin{tabular}{|c|c|c|c|c|}
\hline 
$\left(n_{\rho},n_{z}\right)$ & $\left(0,0\right)$ & $\left(1,0\right)\ \left(0,1\right)$ & $\left(1,1\right)$ $\left(2,0\right)/\left(0,2\right)$ & $\left(2,1\right)/\left(0,3\right)$ $\left(1,2\right)/\left(3,0\right)$\tabularnewline
\hline 
\hline 
$\begin{array}{c}
N\left(l=1\right)\\
\Delta\left(l=3\right)
\end{array}$ & $\begin{array}{c}
940^{+}\\
1232^{+}
\end{array}$ & $\begin{array}{c}
1440\\
1600^{+}
\end{array}$ $\begin{array}{c}
1535^{-}\\
1700^{-}
\end{array}$ & $\begin{array}{c}
1655^{-}\\
1940{}^{-}
\end{array}$ $\begin{array}{c}
1710^{+},?\\
1920^{+},?
\end{array}$ & $\begin{array}{c}
2090_{*}^{-},?\\
?,?
\end{array}$$\begin{array}{c}
2100_{*}^{+},?\\
?,?
\end{array}$\tabularnewline
\hline 
\end{tabular}
\par\end{centering}

\protect\caption{Experimental data of baryon mass for various baryonic states}
\end{table}

\section{Two body baryon interaction}

In this section we will consider the interaction between two baryons
for two-flavor case since it would be more realistic, thus $N_{f},\ k=2$.
For the two-flavor case in the matrix model (\ref{eq:Matrix action}),
integrating out of the $U(k)$- adjoint field $D_{AB}$ gives the
vacuum configuration, which is just the ADHM constraints

\begin{equation}
\vec{\tau}_{\ \dot{\beta}}^{\dot{\alpha}}\left(\bar{X}^{\dot{\beta}\alpha}X_{\dot{\alpha}\alpha}+\bar{\omega}^{\dot{\beta}\alpha}\omega_{\dot{\alpha}\alpha}\right)_{BA}=0\ ,
\end{equation}

\noindent where the indices for baryon are $A,B=1,2...k$. Since our
matrix model is also a deformed ADHM matrix model, we will also use
the ADHM data for our model as done in \cite{key-19 Matrix model for baryons}

\begin{eqnarray}
X^{M} & = & \tau^{3}\frac{r_{M}}{2}+\tau^{1}Y_{M}\ ,\nonumber \\
\omega_{\dot{\alpha}i}^{A=1}=U_{\dot{\alpha}i}^{A=1}\rho_{1} & ; & \omega_{\dot{\alpha}i}^{A=2}=U_{\dot{\alpha}i}^{A=2}\rho_{2}\ ,
\end{eqnarray}

\noindent where $r_{M}$ is the inter-baryon distance, $U^{A}$ is
the $SU\left(2\right)$ matrices which represents the moduli parameters
of each baryon and

\begin{equation}
Y_{M}=-\frac{\rho_{1}\rho_{2}}{4\left(r_{L}\right)^{2}}\mathrm{Tr}\left[\bar{\sigma}_{M}r_{N}\sigma_{N}\left(\left(U^{1}\right)^{\dagger}U^{2}-\left(U^{2}\right)^{\dagger}U^{1}\right)\right]\ ,\label{eq:ADHM data Y_M}
\end{equation}

\noindent we have used $\sigma_{M}=\left(i\vec{\tau},1\right)$ and
$\bar{\sigma}_{M}=\left(-i\vec{\tau},1\right)$. Note that the ADHM
data are just the solution for two Yang-Mills instantons which has
been explicitly used in the approach of soliton \cite{key-17 SS model nuclear force}.
So after the quantization, their degrees of freedom become the spin
and the isospin of each baryon, which are nothing but the gauge rotations
of the flavor gauge group. Here we use real unit vectors $a_{M}^{A}$
to write them as 

\begin{equation}
U^{A}=a_{4}^{A}+ia_{i}^{A}\tau^{i}\ ,
\end{equation}

\noindent with $\left(a_{4}^{A}\right)^{2}+\left(a_{i}^{A}\right)^{2}=1$.
Then we can obtain some useful expression listed as follows which
we are going to use

\begin{eqnarray}
r_{M}Y_{M} & = & 0\ ,\nonumber \\
Y_{M}Y_{M} & = & -\frac{\rho_{1}^{2}\rho_{2}^{2}}{8\left(r_{M}\right)^{2}}\mathrm{Tr}\left[\left(\left(U^{1}\right)^{\dagger}U^{2}-\left(U^{2}\right)^{\dagger}U^{1}\right)^{2}\right]=-\frac{\rho_{1}^{2}\rho_{2}^{2}}{4\left(r_{M}\right)^{2}}\left[1-\left(a_{M}^{1}a_{M}^{2}\right)^{2}\right]\ ,\nonumber \\
\mathrm{Tr}\left[\left(U^{1}\right)^{\dagger}U^{2}\right] & = & \mathrm{Tr}\left[\left(U^{2}\right)^{\dagger}U^{1}\right]\ =\ 2a_{M}^{1}a_{M}^{2}\ .\label{eq:useful formulation}
\end{eqnarray}

Then, we shall compute the two-baryon interaction potential. First
we write down the terms in the action (\ref{eq:Matrix action}) with
ADHM data

\begin{equation}
\mathrm{Tr}\left(D_{0}X^{M}\right)^{2}=2\left[\left(A_{0}^{1}\right)^{2}r_{M}^{2}+\left(A_{0}^{2}\right)^{2}\left(r_{M}^{2}+4Y_{M}^{2}\right)+4\left(A_{0}^{3}\right)^{2}Y_{M}^{2}\right]-8A_{0}^{1}A_{0}^{3}r_{M}Y_{M}\ .\label{eq:DX}
\end{equation}

\noindent Note that the last term in (\ref{eq:DX}) vanishes once
(\ref{eq:useful formulation}) is added. Then, for the kinetic term
of $\omega$ we have

\begin{eqnarray}
\mathrm{Tr}\left(D_{0}\bar{\omega}_{i}^{\dot{\alpha}}D_{0}\omega_{i\dot{\alpha}}\right) & = & 2\left(\rho_{1}^{2}+\rho_{2}^{2}\right)\left[\left(A_{0}^{0}\right)^{2}+\left(A_{0}^{1}\right)^{2}+\left(A_{0}^{2}\right)^{2}+\left(A_{0}^{3}\right)^{2}\right]\nonumber \\
 &  & +4\rho_{1}\rho_{2}A_{0}^{0}A_{0}^{1}\mathrm{Tr}\left[\left(U^{1}\right)^{\dagger}U^{2}\right]+4\left(\rho_{1}^{2}-\rho_{2}^{2}\right)A_{0}^{0}A_{0}^{3}\ .
\end{eqnarray}

\noindent We can minimize it with $A_{0}^{2}=0$ since the component
$A_{0}^{2}$ appears only as $\left(A_{0}^{2}\right)^{2}$. Thus the
resulting baryon interaction potential $V$ from the kinetic term
plus Chern-Simons term can be evaluated by $\int dtV=-S_{\mathrm{on-shell}}$,
which is

\begin{eqnarray}
S_{kinetic}^{\mathrm{on-shell}} & = & \frac{\lambda N_{c}M_{KK}}{54\pi}\left(1+\zeta\right)^{3/2}\mathrm{Tr}\int dt\left[\left(D_{0}X^{M}\right)^{2}+D_{0}\bar{\omega}_{i}^{\dot{\alpha}}D_{0}\omega_{i\dot{\alpha}}\right]+N_{c}\mathrm{Tr}\int dtA_{0}\nonumber \\
 & = & \frac{\lambda N_{c}M_{KK}}{54\pi}\left(1+\zeta\right)^{3/2}\int dt\bigg[2\left(A_{0}^{1}\right)^{2}r_{M}^{2}+8\left(A_{0}^{3}\right)^{2}Y_{M}^{2}\nonumber \\
 &  & +2\left(\rho_{1}^{2}+\rho_{2}^{2}\right)\left(\left(A_{0}^{0}\right)^{2}+\left(A_{0}^{1}\right)^{2}+\left(A_{0}^{3}\right)^{2}\right)\nonumber \\
 &  & +4\rho_{1}\rho_{2}A_{0}^{0}A_{0}^{1}\mathrm{Tr}\left(\left(U^{1}\right)^{\dagger}U^{2}\right)+4\left(\rho_{1}^{2}-\rho_{2}^{2}\right)A_{0}^{0}A_{0}^{3}+\frac{108\pi}{\lambda M_{KK}}\left(1+\zeta\right)^{-3/2}A_{0}\bigg]\ .\label{eq:kinetic onshell}
\end{eqnarray}

\noindent Since the action (\ref{eq:kinetic onshell}) does not depend
on the derivative of $A_{0}$, we can integrate them out straightforwardly,
therefore the resulting baryon potential is

\begin{equation}
V=\frac{27\pi N_{c}}{\lambda M_{KK}}\frac{1}{\left(1+\zeta\right)^{3/2}}\frac{1}{\rho_{1}^{2}\rho_{2}^{2}}\frac{\left(\rho_{1}^{2}+\rho_{2}^{2}+r_{M}^{2}\right)\left(u\rho_{1}^{2}\rho_{2}^{2}-4\left(\rho_{1}^{2}+\rho_{2}^{2}\right)r_{M}^{2}\right)}{\left[u\left(\rho_{1}^{4}-\left(2+u\right)\rho_{1}^{2}\rho_{2}^{2}+\rho_{2}^{4}\right)+5u\left(\rho_{1}^{2}+\rho_{2}^{2}\right)r_{M}^{2}-16\left(r_{M}^{2}\right)^{2}\right]}\ ,\label{eq:potential from kinetic part}
\end{equation}

\noindent where we have defined $u=\left(\mathrm{Tr}\left[\left(U^{1}\right)^{\dagger}U^{2}\right]\right)^{2}-4$.
And there is another part of baryon interaction potential, which comes
from the mass term of $X^{4}$ in action (\ref{eq:Matrix action})
in addition to this potential (\ref{eq:potential from kinetic part}).
It can be evaluated as

\begin{eqnarray}
 & \frac{\lambda N_{c}M_{KK}}{54\pi}\left(1+\zeta\right)^{3/2}\frac{2}{3}\left(1-\frac{1}{2}\zeta\right)M_{KK}^{2}\mathrm{Tr}\left(X^{4}\right)^{2}\nonumber \\
= & \frac{\lambda N_{c}M_{KK}}{81\pi}\left(1+\zeta\right)^{3/2}\left(1-\frac{1}{2}\zeta\right)M_{KK}^{2}\left(\frac{r_{4}^{2}}{2}+Y_{4}^{2}\right) & .\label{eq:mass term X4}
\end{eqnarray}

\noindent This term (\ref{eq:mass term X4}) does not contribute to
the two-baryon interaction since the term of $r_{4}^{2}$ is the mass
term in the single-baryon Hamiltonian. In fact, the off-diagonal elements
of $Y_{4}$ contribute to the interaction between the baryons. Then
using (\ref{eq:ADHM data Y_M}) we have

\begin{equation}
Y_{4}=-\frac{\rho_{1}\rho_{2}}{2r_{M}^{2}}r_{i}\mathrm{Tr}\left(i\tau^{i}U^{(1)\dagger}U^{(2)}\right)\ ,\label{eq:Y4}
\end{equation}

\noindent where $i=1,2,3$, so we can rewrite the potential energy
(\ref{eq:mass term X4}) as

\begin{equation}
\frac{\lambda N_{c}M_{KK}}{162\pi}\left(1+\zeta\right)^{3/2}\left(1-\frac{1}{2}\zeta\right)M_{KK}^{2}\left[r_{4}^{2}+\frac{\rho_{1}^{2}\rho_{2}^{2}}{\left(r_{M}^{2}\right)^{2}}\left(r_{i}\mathrm{Tr}\left[i\tau^{i}U^{(1)\dagger}U^{(2)}\right]\right)^{2}\right]\ .
\end{equation}

\noindent Therefore, putting all together, we obtain the two-body
interaction Hamiltonian of baryons which is

\begin{eqnarray}
V & = & \frac{27\pi N_{c}}{\lambda M_{KK}}\frac{1}{\left(1+\zeta\right)^{3/2}}\frac{1}{\rho_{1}^{2}\rho_{2}^{2}}\frac{\left(\rho_{1}^{2}+\rho_{2}^{2}+r_{M}^{2}\right)\left(u\rho_{1}^{2}\rho_{2}^{2}-4\left(\rho_{1}^{2}+\rho_{2}^{2}\right)r_{M}^{2}\right)}{\left[u\left(\rho_{1}^{4}-\left(2+u\right)\rho_{1}^{2}\rho_{2}^{2}+\rho_{2}^{4}\right)+5u\left(\rho_{1}^{2}+\rho_{2}^{2}\right)r_{M}^{2}-16\left(r_{M}^{2}\right)^{2}\right]}\nonumber \\
 &  & +\frac{\lambda N_{c}M_{KK}}{162\pi}\left(1+\zeta\right)^{3/2}\left(1-\frac{1}{2}\zeta\right)M_{KK}^{2}\left[\frac{\rho_{1}^{2}\rho_{2}^{2}}{\left(r_{M}^{2}\right)^{2}}\left(r_{i}\mathrm{Tr}\left[i\tau^{i}U^{(1)\dagger}U^{(2)}\right]\right)^{2}\right]\nonumber \\
 &  & -\frac{27\pi N_{c}}{4\lambda M_{KK}}\frac{1}{\left(1+\zeta\right)^{3/2}}\left(\frac{1}{\rho_{1}^{2}}+\frac{1}{\rho_{2}^{2}}\right)\ .\label{eq:Baryon potential}
\end{eqnarray}

\noindent If we assume that the size $\rho$ of the baryons or nucleons
is much small, then we can expand (\ref{eq:Baryon potential}) for
$r_{M}\gg\rho$ and obtain a leading term which is

\begin{eqnarray}
V & = & \frac{27\pi N_{c}}{64\lambda M_{KK}}\frac{1}{\left(1+\zeta\right)^{3/2}}\frac{1}{r_{M}^{2}}\left[32+6u+\left(5u+16\right)\left(\frac{\rho_{1}^{2}}{\rho_{1}^{2}}+\frac{\rho_{2}^{2}}{\rho_{1}^{2}}\right)\right]\nonumber \\
 &  & +\frac{\lambda N_{c}M_{KK}^{3}}{162\pi}\left(1+\zeta\right)^{3/2}\left(1-\frac{1}{2}\zeta\right)\left[\frac{\rho_{1}^{2}\rho_{2}^{2}}{\left(r_{M}^{2}\right)^{2}}\left(r_{i}\mathrm{Tr}\left[i\tau^{i}U^{(1)\dagger}U^{(2)}\right]\right)^{2}\right]\ ,\label{eq:Baryon potential r>>rho}
\end{eqnarray}

\noindent again by choosing $\rho_{1}=\rho_{2}=\rho$, we have a formula
from (\ref{eq:Baryon potential r>>rho})

\begin{eqnarray}
V & = & \frac{27\pi N_{c}}{4\lambda M_{KK}}\frac{1}{\left(1+\zeta\right)^{3/2}}\frac{1}{r_{M}^{2}}\left(\mathrm{Tr}\left[U^{(1)\dagger}U^{(2)}\right]\right)^{2}\nonumber \\
 &  & +\frac{\lambda N_{c}M_{KK}^{3}}{162\pi}\left(1+\zeta\right)^{3/2}\left(1-\frac{1}{2}\zeta\right)\left[\frac{\rho^{4}}{\left(r_{M}^{2}\right)^{2}}\left(r_{i}\mathrm{Tr}\left[i\tau^{i}U^{(1)\dagger}U^{(2)}\right]\right)^{2}\right]\ .\label{eq:Baryonic potential final}
\end{eqnarray}

Then, let us compute the vacuum expectation value by using the potential
(\ref{eq:Baryonic potential final}). We use $\left(\vec{I}_{A},\vec{J}_{A},n_{\rho},n_{z}\right)$
to label the states of the two baryons, with $A=1,2$ representing
the two baryons. The excited baryon states are labeled by $n_{\rho}$
and $n_{Z}$ which are the quantum numbers. And $\vec{I},\vec{J}$
is the isospin (spin) of the baryon state respectively. For nucleons
we have $\left(\left|\vec{I}\right|=\left|\vec{J}\right|=\frac{1}{2}\right)$
and the explicit spin/isospin wave functions reads \cite{key-17 SS model nuclear force,key-19 Matrix model for baryons}

\begin{equation}
\frac{1}{\pi}\left(\tau^{2}U\right)_{IJ}=\left(\begin{array}{cc}
|p\uparrow> & |p\downarrow>\\
|n\uparrow> & |n\downarrow>
\end{array}\right).
\end{equation}

\noindent The baryon potential is evaluated as

\begin{eqnarray}
\bigg<\left(\mathrm{Tr}\left[U^{(1)\dagger}U^{(2)}\right]\right)^{2}\bigg>_{I_{1},J_{1},I_{2},J_{2}} & = & 1+\frac{16}{9}\left(\vec{I}_{1}\cdot\vec{I}_{2}\right)\left(\vec{J}_{1}\cdot\vec{J}_{2}\right),\nonumber \\
\bigg<\mathrm{Tr}\left[i\tau^{i}U^{(1)\dagger}U^{(2)}\right]\mathrm{Tr}\left[i\tau^{j}U^{(1)\dagger}U^{(2)}\right]\bigg>_{I_{1},J_{1},I_{2},J_{2}} & = & \delta^{ij}+\frac{16}{9}\vec{I}_{1}\cdot\vec{I}_{2}\left(J_{1}^{i}J_{2}^{j}+J_{2}^{j}J_{1}^{i}-\delta^{ij}\vec{J}_{1}\cdot\vec{J}_{2}\right).\label{eq:formulations for nuclear states}
\end{eqnarray}

\noindent By the standard definition of $S_{12}=12\left(\vec{J}_{1}\cdot\hat{\vec{r}}\right)\left(\vec{J}_{2}\cdot\hat{\vec{r}}\right)-4\vec{J}_{1}\cdot\vec{J}_{2}$
with $\hat{\vec{r}}=\vec{r}/\left|\vec{r}\right|$, we have

\begin{eqnarray}
V_{C}^{(0)}\left(\vec{r}\right) & = & \pi\left[\frac{3^{3}}{2}+8\left(\vec{I}_{1}\cdot\vec{I}_{2}\right)\left(\vec{J}_{1}\cdot\vec{J}_{2}\right)\right]\frac{N_{c}}{\lambda M_{KK}}\frac{1}{\left(1+\zeta\right)^{3/2}}\frac{1}{r^{2}},\nonumber \\
V_{T}^{(0)}\left(\vec{r}\right) & = & 2\pi\left(\vec{I}_{1}\cdot\vec{I}_{2}\right)\frac{N_{c}}{\lambda M_{KK}}\frac{1}{\left(1+\zeta\right)^{3/2}}\frac{1}{r^{2}}.\label{eq:Nuclear potential zero order}
\end{eqnarray}

\noindent For the leading order which comes from the second term in
(\ref{eq:Baryonic potential final}), we have\footnote{Here the exact result is $\frac{\rho^{4}\left(r^{i}\right)^{2}}{\left(r^{M}r^{M}\right)^{2}}$,
we simply write it as $\frac{\rho^{4}}{r^{2}}$.}

\begin{eqnarray}
V_{C}^{(1)}\left(\vec{r}\right) & = & \left[\frac{1}{81}-\frac{16}{2187}\left(\vec{I}_{1}\cdot\vec{I}_{2}\right)\left(\vec{J}_{1}\cdot\vec{J}_{2}\right)\right]\left(1+\zeta\right)^{3/2}\left(1-\frac{1}{2}\zeta\right)\frac{\lambda N_{c}M_{KK}^{3}}{\pi}\frac{\rho^{4}}{r^{2}},\nonumber \\
V_{T}^{(1)}\left(\vec{r}\right) & = & \frac{8}{2187}\left(\vec{I}_{1}\cdot\vec{I}_{2}\right)\left(1+\zeta\right)^{3/2}\left(1-\frac{1}{2}\zeta\right)\frac{\lambda N_{c}M_{KK}^{3}}{\pi}\frac{\rho^{4}}{r^{2}}.\label{eq:Nuclear potential leading order}
\end{eqnarray}

\noindent Thus our total potential up to leading order is 

\begin{eqnarray}
V_{C}\left(\vec{r}\right) & = & V_{C}^{(0)}\left(\vec{r}\right)+V_{C}^{(1)}\left(\vec{r}\right),\nonumber \\
V_{T}\left(\vec{r}\right) & = & V_{T}^{(0)}\left(\vec{r}\right)+V_{T}^{(1)}\left(\vec{r}\right),\nonumber \\
V_{nuclear} & = & V_{C}\left(\vec{r}\right)+S_{12}V_{T}\left(\vec{r}\right).
\end{eqnarray}

This is the short-distance two-body force for baryons in D0-D4/D8
system obtained from our matrix model. With $\zeta<2$ we also find
there is a repulsive core of baryons or nucleons in this system. As
in \cite{key-17 SS model nuclear force,key-19 Matrix model for baryons},
the repulsive potential scales as $r^{-2}$ which was treated as a
property peculiar to the holographic model . If comparing (\ref{eq:Baryonic potential final})
(\ref{eq:Nuclear potential zero order}) (\ref{eq:Nuclear potential leading order})
with the results in \cite{key-19 Matrix model for baryons}, we can
see that the nuclear force in \cite{key-19 Matrix model for baryons}
is just our results with $\zeta=0$ i.e. no smeared D0-branes. With
the appearance of smeared D0-branes, the zeroth-order result is depressed
by the number of D0 while the leading order result increases. As mentioned
in previous sections, our model for baryons is consistent with $\zeta<2$
and this is the condition for the existence of stable baryons in our
theory. However an attractive force appears from (\ref{eq:Nuclear potential leading order})
if $\zeta>2$ which may give an unstable state for baryons in two-body
system by this model.

\section{Summary and discussion}

We have studied the holographic baryons by the matrix model in the
D0-D4/D8 system. We start with considering the baryon vertex inside
the flavor branes i.e. D4'-branes wrapped on $S^{4}$ which is embedded
in the D0-D4 background on large $N_{c}$ limit. And we use the standard
technique in string theory to derive our matrix model. By using T-duality
and dimensional reduction, we obtain our matrix model (\ref{eq:Matrix action})
with $U\left(k\right)$ symmetry, which could also be able to describe
multi-baryons. In order to describe stable baryonic states, we find
the value of $\zeta$ is restricted to $\zeta<2$. This is exactly
the same as the results in \cite{key-24 Baryons in D0-D4}. However
in \cite{key-24 Baryons in D0-D4}, baryons are described by using
BPST configurations which is similar as \cite{key-16 Hata Baryons from instantons}.
In our paper, as a difference, we start from the baryon vertex, but
we come to the same conclusion i.e. stable baryons exist in D0-D4/D8
system only if $\zeta<2$. So it is worthy to believe such a result
is unique for this system.

With this matrix model, we also determine the holographic size and
the spectrum of baryon for the case with $k=1$ and $N_{f}=1,2$.
We find the spectrum obtained from our model could be more close to
the experimental data by choosing suitable value of $\zeta$. Thus
we suggest that it is an effective description of QCD for low energy
baryonic states. Again, we have seen that in the two flavor case the
spectrum of baryon's mass would be totally imaginary if $\zeta>2$.
It turns out that baryons in this system are not stable if $\zeta>2$
as mentioned. In two flavor case, our spectrum of baryon's mass (\ref{eq:Baryon mass})
could recover the results in \cite{key-19 Matrix model for baryons}
if setting $\zeta=0$ i.e. no smeared D0-branes. However, we can not
recover the results (\ref{eq:Hamiltonian in Ref.}) from \cite{key-16 Hata Baryons from instantons},
since the first term in (\ref{eq:Hamiltonian finite Nc two flavor})
is deformed by a factor $4/5$, even if setting $\zeta=0$. This puzzle
was found in \cite{key-19 Matrix model for baryons} first which makes
our spectrum (\ref{eq:Baryon mass}), from the matrix model of D0-D4/D8
system, a bit different from the results by the instanton viewpoint
in \cite{key-24 Baryons in D0-D4}. We are less clear about this and
would take further study about this in the future work. 

And we also compute the two-body short-distance effective potential
for baryons in D0-D4/D8 system i.e. $k=2$ case. It also exhibits
a repulsive core and a tensor force as \cite{key-17 SS model nuclear force,key-19 Matrix model for baryons},
which has been well-known in nuclear physics, but as a difference,
it is modified by the appearance of smeared D0-branes. As it is known
the nucleons interact each others by interchanging mesons and the
effective potential can be derived from the Yukawa coupling. And on
the other hand, in \cite{key-21 SS model in D0-D4,key-24 Baryons in D0-D4},
the spectrum of mesons, such as $\rho$ mesons, of Sakai-Sugimoto
model in D0-D4 background are modified by the appearance of smeared
D0-branes. So in our paper, (\ref{eq:Nuclear potential zero order})
(\ref{eq:Nuclear potential leading order}) can be interpreted as
follows: since the mass of mesons are modified, thus the effective
potential from the interaction are also modified by the appearance
of smeared D0 branes. This two-body interaction has not been calculated
by using the instanton viewpoints in D0-D4/D8 system in \cite{key-24 Baryons in D0-D4},
so our work would be advancing on this front. Our effective two-body
potential (\ref{eq:Nuclear potential zero order}) (\ref{eq:Nuclear potential leading order})
is obtained by expanding (\ref{eq:Baryon potential}) respected to
$\rho/r$, however it shows that if $\zeta>2$, the leading order
potential would be negative i.e. it is an attractive force. Furthermore,
if $\zeta$ and $\lambda$ are large enough, the zero order potential
would be depressed very much while the leading order potential becomes
a rapid negative increasing, which makes the total potential negative.
A negative two-body short-distance effective potential implies there
is an attractive force between baryons or nucleons at short distance,
so the system would not be stable. From the effective two-body force,
it also and consistently turns out that $\zeta$ is restricted even
in two-body system if there would be stable baryonic states.

Finally, we would like to give some more comments about this work.
Basically, our analysis is done in large $N_{c}$ limit as in most
analysis in gauge-gravity duality, though we have tried to give an
effective description of finite $N_{c}$ theory in this paper. The
interaction between baryons would be important for studying on the
phase structure of strong-coupling QCD by holography, such as \cite{key-15 Holographic nuclear physics,key-29 Siwen}.
So a holographic model corresponded to real QCD would be significant,
however unfortunately, currently out of reach. Besides, we can see
that our two-body effective potential for baryons makes sense at short
distance only, and furthermore we still do not know how to introduce
a realistic attractive force consistently or describe the long-distance
force by considering the baryon vertex directly. These problems may
be solved by calculating the interaction between the D4'-branes in
curved spacetime, however it would be a huge challenge in string theory.
So it should be understood as that there is still a long way to the
realistic and strong-coupling QCD from holography.

\section*{Acknowledgments}

This work is inspired by a seminar given by Chao Wu on his work \cite{key-21 SS model in D0-D4,key-24 Baryons in D0-D4}
in USTC. And we would like to thank Qun Wang, Chao Wu and Xiao-Liang
Xia for helpful discussions.

\end{document}